\documentclass[twocolumn,showpacs,nofootinbib,10pt]{revtex4-1}

\usepackage{color}
\usepackage{graphicx}
\usepackage{graphicx,float}\usepackage[all]{xy}
\usepackage{amsmath,upgreek}
\usepackage{amssymb}

\newcommand{\beq}{\begin{eqnarray}}
\newcommand{\eeq}{\end{eqnarray}}
\newcommand{\be}{\begin{equation}}
\newcommand{\ee}{\end{equation}}
\newcommand{\bea}{\begin{eqnarray}}
\newcommand{\eea}{\end{eqnarray}}

\newcommand{\ba}{\begin{eqnarray}}
\newcommand{\ea}{\end{eqnarray}}
\newcommand{\approptoinn}[2]{\mathrel{\vcenter{
  \offinterlineskip\halign{\hfil$##$\cr
    #1\propto\cr\noalign{\kern2pt}#1\sim\cr\noalign{\kern-2pt}}}}}

\newcommand{\appropto}{\mathpalette\approptoinn\relax}
\usepackage[colorlinks,hyperindex]{hyperref}

\begin{document}
\title{
Black hole acoustics in the minimal geometric deformation of a de Laval nozzle
}

\author{Rold\~ao da Rocha}
\affiliation{Centro de Matem\'atica, Computa\c c\~ao e Cogni\c c\~ao, Universidade Federal do ABC - UFABC\\ 09210-580, Santo Andr\'e, Brazil.}
\email{roldao.rocha@ufabc.edu.br}



\begin{abstract} 
The correspondence between sound waves, in a de Laval propelling nozzle, and quasinormal modes emitted by brane-world black holes deformed by a 5D bulk Weyl fluid are here explored and scrutinised. 
The analysis of sound waves patterns in a de Laval nozzle at a laboratory,   reciprocally, is here shown to  provide  relevant data about the  5D bulk Weyl fluid and its on-brane projection, comprised by the minimal geometrically deformed compact stellar distribution on the brane. 
Acoustic perturbations of the gas fluid flow in the de Laval nozzle are proved to coincide to the quasinormal modes of  black holes solutions deformed by the 5D Weyl fluid, in  the geometric deformation procedure. Hence, in a phenomenological E\"otv\"os-Friedmann fluid brane-world model, the realistic shape of a de Laval nozzle is derived and its consequences studied. \end{abstract}



\maketitle

\section{Introduction}

General Relativity (GR) is a well-succeeded theory, widely tested by  experiments and observations, however a limited one to comprise some recent questions, like the dark energy/dark matter. 
GR can be recovered from models involving higher dimensions as a very restricted case \cite{Antoniadis:1998ig,maartens}.
In brane-world models, the brane self-energy density  
manifest as the brane tension ($\sigma$), which is assumed to be infinite in the GR limit. Nevertheless, a finite value for the brane tension, in codimension one models, is an ubiquitous  feature of the brane that also comprises the warped five-dimensional (5D) geometry, besides the brane self-gravity. 

Although an enormous brane tension value recovers GR at low energy regimes,  phenomenological evidence indicates a variable brane tension \cite{Gergely:2008jr,Gergely:2008fw,Kanno:2003xy,Abdalla:2009pg}.  At an immensely hot universe, the tension of the brane had attained a nugatory value. Afterwards, the brane tension increased  and the universe cooled down \cite{Gergely:2008jr,Gergely:2008fw,Kanno:2003xy}. Fluid membranes play a central role in modelling this scenario, wherein  
a temperature-dependent tension is ruled by the E\"otv\"os principle \cite{Abdalla:2009pg}, that governs a Friedmann brane with a  scale factor that drives the expansion of the universe  \cite{Gergely:2003pn}.

4D gravity  can be  effectively formulated on a brane-world by 
two complementary methods. The first one comprises 
the  Shiromizu-Maeda-Sasaki implementation of the Gauss-Codazzi on-brane projection routine \cite{GCGR}. Nevertheless, this method does not represent a consistent system of equations, since the 5D bulk Weyl tensor can not be  determined from data on the brane. In fact, there is no action, whose projected Euler-Lagrange equations  onto the brane can be derived \cite{maartens}. 
A complementary technique does involve an action that, at low energy regimes, derives the effective 4D theory described by the respective Euler-Lagrange equations \cite{Nojiri:2000zx}. Hence, these two complementary procedures can  be employed  in the construction of an effective 4D theory \cite{Kanno:2003xy}.

Among successful efforts to formulate theories beyond GR, the procedure consisting in accomplishing a minimal geometric deformation of the Schwarzschild solution in a brane-world was derived \cite{ovalle2007,Ovalle:2007bn,covalle2}. It comprises exact solutions of the Einstein's  equations for the 4D effective theory on the brane \cite{ovalle2007,Ovalle:2007bn,covalle2}. The minimal geometric deformation procedure incorporates high-energy improvements to GR, when the (brane) vacuum state is percolated by a  5D Weyl fluid in the bulk  \cite{ovalle2007,Ovalle:2013xla,darkstars}.  These analytical solutions -- of the 4D brane effective Einstein's field equations --  encode compact distributions supporting stellar systems that can even exhibit solid crusts \cite{darkstars,Ovalle:2013xla}, driven by the 5D bulk Weyl fluid, and also peculiar generalizations \cite{Ovalle:2015nfa}.   The deformation itself comprises the brane tension as the  managing parameter of high energy regimes. This setup has GR corresponding to an ideally rigid brane ($\sigma\to\infty$), at low energies. A more refined 
setup can be implemented by considering a variable tension fluid brane \cite{Gergely:2008jr,Casadio:2013uma}. 
This approach has been  comprehensively and successfully constrained by experimental and observational bounds, provided by the perihelion precession of Mercury, 
  the deflection of light by the Sun,  the gravitational redshift and the radar echo delay, recently obtained in Ref.  \cite{Casadio:2015jva}. 
Besides, observational lensing effects by the minimal geometrically  deformed black holes have a  typical signature that may be soon probed by  the European Space Agency  satellite mission \cite{Cavalcanti:2016mbe}. Despite of this comprehensive list of experimental and observational possible signatures, regarding the brane-world black hole that is geometrically deformed by a 5D Weyl fluid, any study of the quasinormal modes produced by this kind of black holes lacks, still. 

On the other hand, acoustic perturbations of a gas flow, in the so-called de Laval nozzle, have been  shown to correspond to the general form of perturbations of Schwarzschild black holes \cite{Okuzumi:2007hf,Abdalla:2007dz,Cuyubamba:2013iwa}. It introduces  
the feasibility to produce and observe their quasinormal
resonances  in a laboratory. 
The de Laval nozzle is an example among  
propelling nozzles, that are widely studied devices that  turn a fluid (gas)  turbine into a jet engine. A de Laval nozzle is constituted by an hourglass-shaped tube, strained in the middle, utilized to accelerate a hot pressurised gas to a high supersonic speed into the thrust direction. The fluid thermal energy is commuted into kinetic energy, and the fluid velocity increases.  The energy to accelerate the gas stream induces the  gas to adiabatically expand with high efficiency to a final -- transonic, supersonic, or even hypersonic speed --  propelling jet. 
The de Laval nozzle is constructed upon the theory of quasi-1D flows, where a fluid moves at the magnitude of the speed of sound. In this regime, the changes in the fluid density turns significant and compress the  flow. The de Laval nozzle is based upon the Venturi effect. 

Fluid flows have been studied, in this context, in a de Laval nozzle, aiming to observe  acoustic black holes  in the Schwarzschild setup \cite{Okuzumi:2007hf,Abdalla:2007dz,Cuyubamba:2013iwa}. The acoustic black hole surface gravity was experimentally derived in a laboratory, in Ref. \cite{Furuhashi:2006dh}.
As argued in Refs. \cite{Abdalla:2007dz,Cuyubamba:2013iwa}, sonic regions in a fluid can induce a surface for the sound waves, known as the acoustic horizon, that emulates a black hole event horizon. Perturbations of sound waves have been shown to be analogue to the quasinormal modes, corresponding to black holes gravitational excitations  
\cite{Kokkotas:1999bd,Nollert:1999ji}, being moreover successfully explored in different contexts \cite{Konoplya:2011qq,Chirenti:2012ap,Chirenti:2012ap,Morgan:2009vg}. 

Our point here is to derive and analyse   
the correction to a de Laval nozzle trend, using also its analogy to a brane-world black hole in the 
minimal geometric deformation setup, regarding a variable brane tension. Besides, another goal here is to study the analogy between waves in a de Laval nozzle in a laboratory and quasinormal modes of minimal geometrically deformed brane-world black holes. Hence, to scrutinise sonic waves in a de Laval nozzle can circumvent the indeterminacy of the 5D Weyl tensor on the bulk \cite{maartens,CoimbraAraujo:2005es}  that encrypts the bulk geometry. 
Since the minimal geometric deformation is generated by a 5D Weyl fluid on the bulk, experiments regarding a de Laval nozzle  in a laboratory 
may reciprocally provide relevant data about the 5D bulk  Weyl fluid.

This paper is organized as follows: Sect. II is devoted to a brief review, regarding minimal geometrically deformed compact systems, further refined by phenomenological E\"otv\"os-Friedmann fluid branes. In Sect. III, a gas flow is perturbed in a de Laval nozzle, whose  wave equation is analogue to the wave equation regarding spin-$s$ perturbations of  minimal geometrically deformed brane-world black holes. Hence, the current bound for the variable brane tension provides corrections to the expression for the trend of de Laval nozzles. 
Moreover, quasinormal modes from these black holes are here proposed to be 
studied in a laboratory, when the wave equation in a de Laval nozzle equals the wave equation of spin-$s$ perturbations of brane-world black holes undergoing a minimal geometric deformation. 
Sect. IV is dedicated to discuss our results, to summarise the conclusions and to provide relevant perspectives.

\vspace*{-0.5cm}
\section{The minimal geometric deformation setup and fluid branes}
\label{MGD}
\vspace*{-0.2cm}

Acoustic analogues of brane-world black holes have been reported \cite{Ge:2007ts}, in the context of both Randall-Sundrum 
and Dvali-Gabadadze-Porrati \cite{Randall:1999ee,Dvali:2000hr} setups. However,  no approach has regarded realistic data on the brane yet, comprising the variable brane tension paradigm on fluid branes.  
Employing the minimal geometric deformation technique 
incorporates high-energy refinements to general relativity, by permeating the brane vacuum outer of a compact distribution with a  Weyl fluid in the 5D bulk  \cite{ovalle2007,Ovalle:2013xla,darkstars}. 
Brane-world 
models encompassing a variable brane tension are best implemented by  E\"otv\"os-Friedmann fluid branes, where the brane temperature drives the brane tension, according to the E\"otv\"os' rule across the universe expansion   \cite{Gergely:2008jr,Abdalla:2009pg}.

The most stringent  brane tension bound  $\sigma \gtrapprox  3.2\times10^{-6} \;{\rm GeV^4}$ has been provided in the context of the minimal geometric deformation of black holes, formed as Bose-Einstein condensates  of gravitons that weakly interact among themselves  \cite{Casadio:2016aum}. The associated entropic information content can also predict   
the Chandrasekhar critical density of compact objects in this paradigm \cite{Casadio:2016aum,Gleiser:2015rwa}. 

The effective 4D Einstein's equations can be derived by the Gauss-Codazzi on-brane projection method, from the 5D bulk equations.  \cite{GCGR,maartens}. 
In natural units, the  4D Einstein's effective equations were obtained in Ref. \cite{GCGR} (hereon Greek indexes run in the set of Minkowski space-time indexes):
\begin{equation}
\label{5d4d}
G_{\mu\nu}+\Lambda g_{\mu\nu}
-{\rm T}_{\mu\nu}=0,
\end{equation} where $G_{\mu\nu}$ denotes the Einstein's tensor and the cosmological parameter on the brane is denoted by $\Lambda$. 
The effective stress-energy tensor 
$
{\rm T}_{\mu\nu}
=
T_{\mu\nu}+\sigma^{-1}S_{\mu\nu}+\mathcal{E}_{\mu\nu}+ P_{\mu\nu}+L_{\mu\nu}
$ encrypts the matter stress-energy tensor on the brane ($T_{\mu\nu}$),  the 5D bulk Weyl tensor electric projection on the brane ($\mathcal{E}_{\mu\nu}$) -- that comprise data (constituting a Weyl fluid) about  the gravitational field out of the brane -- and $S_{\mu\nu}$ is the  traceless irreducible component, proportional to the brane extrinsic curvature \cite{GCGR} -- regards 
5D effects onto the brane; the tensor components $L_{\mu\nu}$ encrypt the asymmetric embedding of the
brane, into the bulk, and $P_{\mu\nu}$ stands for the pullback onto the brane of the stress-energy tensor, designating eventual 5D 
non-standard model fields, comprising radiation of quantum origin, dilatonic, and even moduli fields \cite{GCGR,maartens,Gergely:2003pn}. 
Deviations from the usual  Einstein's standard equations in GR can be also generated by excitations of 5D gravitons, whose effects are encompassed in the $P_{\mu\nu}$ tensor.   Eventually, some of the terms constituting $
{\rm T}_{\mu\nu}$ can equal zero.

Exact solutions of the 4D effective Einstein's equations are rare, due to 
the intricacy of the system of equations and to initial data out of the brane as well. Compact distributions, modelling stellar structures are spherically symmetric, static, solutions of Eq. (\ref{5d4d}), of type \begin{equation}\label{abr}
ds^{2} = -A(r) dt^{2} + (B(r))^{-1}dr^{2} + r^{2}d\Omega^2,
\end{equation} 
for $d\Omega^2$ representing the 2-sphere surface element. 
The deformation
on the radial component in Eq. (\ref{abr}) is caused  by the bulk constituents, encrypting not only anisotropic effects originated from the bulk gravity but also the effects of a 5D Weyl fluid in the bulk, 
whose brand permeates the brane vacuum.

The minimal geometric deformation is implemented by  fixing   
the  temporal component in (\ref{abr}) 
and deforming  the outer radial component  \cite{covalle2,darkstars},  
\begin{eqnarray}
\label{deff}
B(r)
=
{1-\frac{2\,M}{r}}
+\varsigma\,e^{\Theta(r)}\ ,
\label{I}
\end{eqnarray}
where 
\begin{eqnarray}\Theta(r)=
\int^r_{{\rm R}}
\frac{f_1(A({\rm a}))}{f_2(A({\rm a}))}\,d {\rm a},\end{eqnarray}
 for \cite{covalle2}
 \begin{subequations}
 \begin{eqnarray}
 f_1(A)&=& \frac{\ln(A)^{\prime}}{r^2}\left(\ln(A)^{\prime}+\frac{2}{r}\right)+\frac{AA''}{A'^{2}}-1\\ 
f_2(A)&=&\left(\frac{2}{r}+\frac12\ln(A)^{\prime}\right)^{-1}.\end{eqnarray}\end{subequations} The prime denotes the derivative with respect to the radial coordinate, and  ${\rm R}$ denotes the compact distribution effective radius \cite{ovalle2007}. The  $\varsigma$ parameter in (\ref{deff}) regards the Weyl fluid in the bulk and its induced deformation of the brane 4D vacuum \cite{Casadio:2015jva}. 
The region inner to the stellar distribution is  regular at the origin.
The inner and outer region to a star have a shared boundary
constituted of a solid incrustation. The (variable) brane tension and the stellar effective radius are parameters that determine the star crust width  \cite{darkstars,Ovalle:2013xla}. 
The outer region $r>{\rm R}$~\cite{ovalle2007} then promotes the  deformed metric ~\cite{covalle2}
\begin{subequations}
\ba
\label{nu}
\!\!\!\!\!\!A(r)
&=&
1-\frac{2\,M}{r}
\ ,
\\
\!\!\!\!\!\!B(r)
&=&
\left[1+\frac{{\varsigma}\mathfrak{l}}{r\left(1-\frac{3\,M}{2\,r}\right)}\,\right]\left(1-\frac{2\,M}{r}\right)
\ ,
\label{mu}
\ea
\end{subequations} 
where \cite{covalle2}
$
\mathfrak{l}
\equiv
\left(1-\frac{2M}{{\rm R}}\right)^{\!-1}\!
\left(1-\frac{3M}{2{\rm R}}\right){\rm R}.
$ 
Refs. \cite{ovalle2007,Casadio:2013uma} show that the metric radial component (\ref{nu}) can be split as 
-- hereon in this section the subindex ``0'' refers to the GR limit $\sigma\to\infty$ (or, equivalently, $\varsigma=0$):
\begin{eqnarray}\label{B2B}
B(r)=B_0(r) + B_\varsigma(r)
\ ,
\end{eqnarray}
where 
\begin{subequations}
\begin{eqnarray}\label{B3B}
\!\!\!\!\!\!\!\!\!\!\!\!\!\!\!\!B_0(r)&\!=\!&\lim_{\varsigma\to0}B(r)=1\!-\!\frac{2M}{r},\\
\!\!\!\!\!\!\!\!\!\!\!\!\!\!\!\!B_\varsigma(r)&=&-\frac{\left({1\!-\!\frac{2M_0}{r}}\right)\frac{1}{r}}{{1\!-\!\frac{3M_0}{2\,r}}}\;\left(\mathfrak{l}\vert_{M_0}\right)\,\varsigma\,,\label{B31B}
\end{eqnarray}
\end{subequations}
where $B_\varsigma(r)$ evinces a high energy correction to the Schwarzschild solution to order ${\cal O}(1/\sigma^{2})$, for  $M=M_0+{\cal O}(1/\sigma)$.

The parameter $\varsigma$ is proportional to 
the stellar distribution compactness and drives the geometric deformation of the Schwarzschild solution, having the 
explicit expression in terms of the brane tension  \cite{ovalle2007,covalle2}:
\begin{eqnarray}
\label{betafinal2}
\!\varsigma
\!&\approx&\!
\frac{\upalpha\,\tau({\rm R})}{98\pi^2\sigma}\,
\frac{\left(\mathfrak{l}\vert_{M_0}\right)}{{\rm R}}
\left[63\upalpha\!+\!390\upalpha^2-\frac{9167}{7}\upalpha^3\!+\!\frac{135952}{63}\upalpha^4\right]\nonumber\\&\appropto&
\frac{\sigma^{-1}}{{\rm R^2}}{\left(\mathfrak{l}\vert_{M_0}\right)}\equiv -{d_0}\frac{\sigma^{-1}}{{\rm R}}
\,,
\end{eqnarray}
for $\upalpha\equiv d_0{\rm R}^2$ and \begin{eqnarray}\tau(r)\equiv \left(1+\upalpha\left(\frac{r}{{\rm R}}\right)^2\right)^{-3}\left(1+3\upalpha\left(\frac{r}{{\rm R}}\right)^2\right)^{-1}.\end{eqnarray} Typically $d_0\approxeq \frac{0.27492}{{\rm R}^2}$. 
The GR $\sigma\to\infty$ limit yields $\varsigma=0$ in Eqs. (\ref{nu}, \ref{mu}), leading to the standard Schwarzschild metric solution of Einstein's equations.

The current experimental and observational data was shown to enforce  the strongest bound $
|\varsigma|\lesssim 6.1 \times 10^{-11}$ (obtained by the 
perihelion precession classical test of GR) and the weakest bound 
$|\varsigma|\lesssim 8.2 \times 10^{-5}$ (derived from the 
radar echo delay classical test of GR)
on the adimensional deformation parameter, in Ref. \cite{Casadio:2015jva}.   
Besides, the most recent brane tension bound $\sigma  \gtrapprox  3.2\times10^{-6} \;{\rm GeV^4}$ has  been obtained by the informational entropy of the minimal geometrically deformed Bose-Einstein condensate of gravitons \cite{Casadio:2016aum}.
Eq. (\ref{betafinal2}) implies a negative value for the parameter $\varsigma$. Hence, the gravitational field strength is mitigated by the finite value of the  brane tension and by the 5D encompassing scenario \cite{Casadio:2015jva}, which attains a maximum at the stellar surface
$r={\rm R}$.  Denoting the stellar distribution density by $\rho_{\rm star}$, the bound $\sigma  \gtrapprox  3.2\times10^{-6} \;{\rm GeV^4}$  
complies with the condition $\frac{\rho_{\rm star}}{\sigma}\ll1$ \cite{Gergely:2008jr,maartens}. 

E\"otv\"os-Friedmann fluid branes are known to have a variable brane tension proportional  to the universe temperature $T$. The E\"otv\"os law 
asserts that  $\sigma\approx T-\tau$, \cite{Gergely:2008jr,Wong:2010rg}, for $\tau$ a crucial constant parameter that drives $\sigma$ into positive values, subsequently to the Big Bang  \cite{Gergely:2008jr,Abdalla:2009pg}.  The scale factor constant value $a_{0}$ fixes the beginning of the universe at a  $\tau$ temperature  \cite{Gergely:2008jr,Wong:2010rg}. Ref. \cite{Gergely:2008jr} computed the temperature dependence upon the scale factor as $T(t) \approx \frac{1}{a(t)}$ \cite{Gergely:2008jr}, then yielding a variable brane tension that is time dependent \cite{Gergely:2008jr,Gergely:2008fw} \begin{eqnarray}\label{tensao}\frac{\sigma(t)}{\sigma_{\rm 0}}=\frac{\kappa^2}{\kappa_0^2}=1-\frac{a_{0}}{a(t)},\end{eqnarray}
where $\kappa^2$ denotes the 4D coupling ``constant'' and $\kappa_0^2=(8\pi G)^{-1}$ is the late-time coupling constant, for $G$ denoting the Newton's constant.

 At early times, until the radiation density had equated the brane matter density, the brane tension could be taken as having a slight value. However, the brane tension and the 4D coupling parameter likewise magnified as the universe expanded. 
The time-dependent brane tension expression yields $\Lambda_{{\rm 4D}}=\Lambda_{\rm 0}-\frac{a_0}{a(t)}\left(1-\frac{a_0}{2a(t)}\right)\kappa^2_{\rm 0}\sigma_{\rm 0}$ \cite{Gergely:2008jr,Gergely:2008fw}. 
 Black string and black brane solutions in a variable brane-world context were studied in Refs. \cite{Casadio:2013uma,Bazeia:2014tua,daRocha:2013ki}.

\section{The minimal geometric deformation of a de Laval nozzle}

A de Laval nozzle is deeply based on the Venturi effect. When the flow of a (gas) fluid  passes through a constricted part of a tube with variable cross section $A(x)$, it originates the reduction in the fluid pressure, whereas the fluid velocity increases. Modelling de Laval nozzles considers quasi-1D flows, which are isentropic, adiabatic, and frictionless ones, which shall be considered hereon. 
 The regarded fluid can consist of an ideal gas and expressed by the equation of state $p=\rho R T$, where $p$ denotes 
     the fluid pressure, $T$ is the absolute temperature, and $R$ is the universal gas constant. An ideal gas is known to have 
     a constant heat capacity, at constant pressure and constant volume -- respectively denoted by $C_p$ and $C_V$. Hence $R = C_p - C_V$ and the  specific heat ratio reads $\gamma= C_p/C_V$. For instance, the heat capacity ratio for Helium is $\gamma = 1.66$, whereas Nitrogen has $\gamma = 1.4$. Hereupon diatomic gas molecules shall be regarded. 
     
     An isentropic gas flow, from an initial to a final  state has the property \cite{landau}
  \begin{eqnarray}\label{isen}
     p = \rho^\gamma=T^{\frac{\gamma}{\gamma-1}}\,,
     \end{eqnarray}      
where, hereon, these quantities shall be normalised by the initial state. 
  Prominent properties of isentropic flows comprise the uniform expansion of the gas, composing then a  shock-free, continuous, flow.
A relevant parameter of a compressible flow is the Mach number, 
$
\mathfrak{M}(x)= \frac{v(x)}{c_s(x)},
$ where $c_s^2=\frac{dp}{d\rho}=\gamma RT$  is the local speed of sound and $x$ denotes the transversal nozzle coordinate, namely, the coordinate along the de Laval nozzle, and $v$ is the local flow velocity.
The Mach number is employed to categorise the distinct regimes of flow\footnote{Those regimes include  hypersonic, supersonic, transonic, sonic, and subsonic flows.}.  
Besides, the mass flow rate $\frac{dm}{dt}$ is the flux per unit throat area $\rho Av$, meaning the mass of gas that  passes through a cross section of the tube per unit time, also known as the fluid discharge \cite{landau}. A quasi-1D fluid flow is ruled by the Euler-Lagrange equations and  the continuity relation in fluid mechanics, given by \cite{landau} 
 \begin{subequations}
\begin{eqnarray}
\frac{\partial}{\partial t}(\rho A) + \frac{\partial}{\partial x}(\rho Av) &\!=\!& 0 \,,
\label{sw0} \\
\frac{\partial}{\partial t}(\rho Av) + \frac{\partial}{\partial x}[(\rho v^2 + p)A] &\!=\!& 0 \,,\label{sw1}
\\
\!\!\!\!\!\!\!\!\!\!\!\!\!\!\!\!\!\!\!\frac{\partial}{\partial t}\!\left(\!\frac{\rho v^2}{2} \!-\! \frac{p}{1\!-\!\gamma}A\right) \!+\!  \frac{\partial}{\partial x}\!\!\left[\left(\frac{\rho v^2}{2} \!-\! \frac{\gamma}{1\!-\!\gamma}\right)Av\right] &\!=\!& 0 \,.\label{sw}
\end{eqnarray}
\end{subequations}
Instead of Eq.~\eqref{sw1}, one can use the Euler's equation\begin{gather}
\rho\left(\frac{\partial v}{\partial t} + v \frac{\partial v}{\partial x}\right)+\frac{\partial p}{\partial x} =0\,,
\label{eueu}
\end{gather}
which can be led to the
Bernoulli's equation  
\beq\label{enta}
\frac{1}{2}\left(\frac{\partial \Phi}{\partial x}\right)^2 + \int\rho^{-1}dp =-\frac{\partial \Phi}{\partial t}  \,,
\label{bno}
\eeq
where the last term in Eq. (\ref{enta}) represents the heat function of a barotropic fluid, identified to the enthalpy, and $\Phi = \int v\,dx$ denoting the velocity potential. From Eq. (\ref{enta}), the linearized equation 
for sound waves can be then obtained, considering perturbations $\phi$ and $\delta\rho$, respectively around $\Phi$ and $\rho$ \cite{Okuzumi:2007hf,Abdalla:2007dz}. 

A quasi-1D fluid flow in a de Laval nozzle has a stagnation state, which is a state attained by the fluid if it is conveyed to rest into an  isentropic state and without work. The stagnation speed of sound is denoted by $c_{s0}$. 
Now, the 
acoustic analogue of the tortoise coordinate, $x^{\star}$, is defined by  
\begin{equation}
x^{\star} = c_{s0} \int \left[{c_{s}(x) (1-\mathfrak{M}(x)^{2})}\right]^{-1}\,dx.
\end{equation} 
Perturbing the system of equations (\ref{sw0} -- \ref{sw}) in a
nozzle yields  \cite{Okuzumi:2007hf}:
\begin{eqnarray}
\biggl[ \frac{d^2}{dx_\star^{2}} + \frac{\omega^2}{c_{s0}^2} - V(x_\star) \biggr] \upphi(\omega,x_\star) = 0, \label{seq1}\end{eqnarray}
where the associated potential, representing the sound waves curvature scattering on the acoustic black hole, reads  
\begin{eqnarray}\label{vxx}
V(x_\star) = \frac{1}{2{\rm g}^2}\left({\rm g}\frac{d^2{\rm g}}{dx^{\star2}}
    - \left(\frac{d{\rm g}}{\sqrt{2}dx_\star}\right)^2\right),
\end{eqnarray}
for \cite{Okuzumi:2007hf,Abdalla:2007dz}
\begin{eqnarray}\label{isen1}
{\rm g}(x) &\equiv&\frac{\rho(x) A(x)}{c_s(x)}\appropto\frac{A(x)}{2\rho^{(\gamma-3)/2}}, \\\label{isen11}
\upphi(\omega,x_\star) &=& \int \sqrt{{\rm g}(x_\star)} \phi(t,x_\star)\,e^{i\omega[t-f(x_\star)]}\,dt,
\end{eqnarray}
where $f(x)$ in Eq. (\ref{isen11}) is a function defined by $\frac{df(x)}{dx}=\frac{|v|}{c_s^2-v^2}$. 

A de Laval nozzle trend is constructed upon the nozzle throat cross-sectional area. Dimensionless quantities for $\rho(x)$ and $A(x)$ are obtained by
measuring them in units of the stagnation gas density $\rho_0$ and of the throat nozzle cross-sectional area, respectively. Moreover, \cite{Abdalla:2007dz,Cuyubamba:2013iwa}
\begin{eqnarray}\label{isen2}
A&\appropto&\left(1-\rho^{(\gamma-1)}\right)^{1/2}\rho,
\end{eqnarray} which by Eq. (\ref{isen1}) yields
$
{\rm g}=\frac{\rho^{1-\gamma}}{2\left(\rho^{1-\gamma}-1\right)^{1/2}}
$, following that 
\begin{eqnarray}
\rho^{1-\gamma}&=&2{\rm g}^2-2{\rm g}\sqrt{{\rm g}^{2}-1}
\label{isen3}
=1+\frac{\gamma-1}{2}\mathfrak{M}^2 \geq 1,
\end{eqnarray} yielding
\begin{equation}
\mathfrak{M}^2=\frac{2}{\gamma-1}(2{\rm g}^2-2{\rm g}\sqrt{{\rm g}^{2}-1}-1).
\end{equation}
The Mach number equals the unit at the horizon, wherein thus ${\rm g}$ have to be finite,
\begin{eqnarray}\label{condt}
{\rm g}_{\rm horizon}=\frac{1+\gamma}{2\sqrt{2\gamma-2}}=\frac{3\sqrt{5}}{5}\geq 1.\end{eqnarray} 
Replacing Eq. (\ref{isen3}) into (\ref{isen1}) implies the 
cross-sectional nozzle area expressed in terms of ${\rm g}$  \cite{Okuzumi:2007hf}, 
\begin{eqnarray}\label{2}
A&=&2^{\frac{1+\gamma}{2\gamma-2}}{\rm g}^{\frac{4-\gamma}{2\gamma-2}}\left({\rm g}-\sqrt{{\rm g}^2-1}\right)^{\frac{2-\gamma}{\gamma-1}},
\\
&=&\frac{1}{\mathfrak{M}^2}\left[\left(1+\frac{\gamma-1}{2}\mathfrak{M}^2\right)\frac{2}{\gamma+1}\right]^{\frac{1+\gamma}{\gamma-1}}\,.\label{cross1}
\end{eqnarray}
On the other hand, the analogy between fluid flows in
a de Laval nozzle and the brane-world black hole --  
undergoing a minimal geometric deformation -- can be implemented. In fact, 
scalar field
perturbations in the minimally-geometric deformed brane-world black hole  background are known to yield the wave-like equation \cite{Konoplya:2011qq} 
\begin{equation}\label{seq2}
\left(\frac{d^2}{dr_*^2}+\omega^2-V(r_*)\right)\Psi(r_*)=0,
\end{equation}
where $dr_*=\frac{dr}{B(r)}$ and the effective potential for 
the quasinormal ringing of the brane-world black hole under 
the minimal geometric deformation reads 
\begin{widetext}
\begin{equation}
\!\!\!\!V(r)\!=\!\left(1\!+\!\frac{{\varsigma}\mathfrak{l}}{r\left(1\!-\!\frac{3\,M}{2\,r}\right)}\right)\left(1\!-\!\frac{2\,M}{r}\right)\Bigg\{\frac{\ell(\ell\!+\!1)}{r^2}\!+\!(1-s^2)\left[{2M}\left(1\!+\!\frac{{\varsigma}\mathfrak{l}}{r\left(1\!-\!\frac{3\,M}{2\,r}\right)}\right)\!-\!\frac{{\varsigma}\mathfrak{l}}{r^2\left(1\!-\!\frac{3\,M}{2\,r}\right)^2}\left(1\!-\!\frac{2\,M}{r}\right)\right]\frac1{r^3}\Bigg\}.
\end{equation}
\end{widetext}
Eq. (\ref{seq2}) is analogue to Eq. (\ref{seq1}). In fact, to find a scalar function  ${\rm g}$ that produces the same potential, the
tortoise coordinate of the black hole solution is identified to the de Laval
nozzle, $dr_*=dx_\star$, yielding 
\begin{eqnarray}
dx_\star^2&=&\frac{\rho^{1-\gamma}}{(1\!-\!\mathfrak{M}^2)^2}dx^2\nonumber\\&=&
\frac{{2{\rm g}^2\!-\!2{\rm g}\sqrt{{\rm g}^{2}\!-\!1}\!-\!1}}{\left[1\!-\!\frac{2}{\gamma-1}
\left(2{\rm g}^2\!-\!2{\rm g}\sqrt{{\rm g}^{2}\!-\!1}\!-\!1\right)\!-\!1\right]^2}dx^2.\label{dxxx}
\end{eqnarray}
The differential equation for $g(r)$ then reads
\begin{equation}\label{eq123}
\!\!\!\!\!\!\left[B(r){\rm g}'(r)\right]^\prime\!-\!B(r)B'(r){\rm g}'\!(r)\!-\!\frac{\left(B(r){\rm g}^\prime(r)\right)^2}{2{\rm g}(r)}\!=\!V(r){\rm g}(r).
\end{equation}
One can elect an unit event horizon radius, yielding the nozzle
coordinate to be written with respect to the event
horizon.

In the limit $\varsigma\to0$, Eq. (\ref{B2B}) is reduced to Eq. (30) in Ref. \cite{Abdalla:2007dz}, which has the solution in Eq. (31) in that reference. Substituting Eq. (\ref{B2B}) into (\ref{eq123}) yields an  intricate equation that can not be analytically solved. However, by splitting the solution  of (\ref{eq123})  into a sum of a purely GR component  (${\rm g}_0(r)\equiv\lim_{\sigma\to\infty}{\rm g}(r)$) and a component that is induced by the 5D Weyl fluid, 
\begin{eqnarray}{\rm g}(r)={\rm g}_0(r) + {\rm g}_\varsigma(r),\label{B5B}\end{eqnarray} we can  substitute the solution of (\ref{eq123}) 
for $\varsigma=0$,  
\begin{eqnarray}\nonumber\label{B4B}
\!{\rm g}_0(r)\!=\!{\frac{1+\gamma}{2\sqrt{2\gamma\!-\!2}}}\sum_{j=s}^{\ell}\!
\left(\frac{ (\ell \!+\! j)!}{( j \!-\! s )! (s\!+\!j )! (\ell \!-\! j)!}r^{j+1}\!\right)^{\!2}\!,
\end{eqnarray} obtained in Ref. \cite{Abdalla:2007dz},  to find 
the ${\rm g}_\varsigma(r)$ function, iteratively solving Eq. (\ref{eq123}). The solution of 
Eq. (\ref{eq123}) has two integration constants, determined by Eq. (\ref{condt}). 
Eq. (\ref{dxxx}) and Eq. (\ref{mu}) then provide the 
nozzle coordinate $x$ with respect to $r$, 
\begin{equation}\label{b6b}
\!\!\!\!\!\!x\!=\!\!\!\intop_{r_0}^r\!\!\frac{2\left[2{\rm g}^2(\bar{r})\!-\!2{\rm g}(\bar{r})\sqrt{{\rm g}^{2}(\bar{r})\!-\!1}\!-\!1)\right]\!-\!1\!-\!\gamma}
{(1\!-\!\gamma)B(\bar{r})\!\left(2{\rm g}^2(\bar{r})\!-\!2{\rm g}(\bar{r})\sqrt{{\rm g}^{2}(\bar{r})\!-\!1}\!-\!1\!\right)^{\!\!1/2}}\,d\bar{r}.
\end{equation}
The integral lower limit  can be made consistent with the fact that the coordinate $x$ is null at the sonic point, by imposing $r_0=1$.

Hence, the de Laval nozzle cross-section $A(x)$ can be finally derived, modelling the nozzle shape. In fact, Eq. (\ref{B4B}) 
can be  input into  Eq. (\ref{eq123}), whose numerical solutions for ${\rm g}(r)$ derive the corrections (due to the 5D Weyl fluid) ${\rm g}_\varsigma(r)$, in Eq. (\ref{B5B}). Subsequently, we rewrite the numerical solution ${\rm g}(r)$ in Eq. (\ref{B5B}) with respect to 
the transversal nozzle coordinate $x$ in Eq. (\ref{b6b}), substituting it into the expression for the nozzle cross section $A(x)$ in Eq. (\ref{cross1}).   

In what follows, the solid gray areas in Figs. 1 - 3 indicate the cross section 
$A(x)$ of the de Laval nozzle and its shape, in the $\sigma\to\infty$ GR limit (gray area limited by the continuous gray line) and its minimal geometric deformation due to a 5D bulk Weyl fluid (gray area limited by the dotted gray line). The black strips respectively represent the 
effective potential for those both cases. The brane tension bound adopted  $\sigma \approx  3.2\times10^{-6} \;{\rm GeV^4}$ was derived in Ref. \cite{Casadio:2016aum} throughout the numerical computation of the minimal geometric deformation case, as well as the weakest experimental bound 
$|\varsigma|\lesssim 8.2 \times 10^{-5}$ 
on the minimal geometric deformation parameter \cite{Casadio:2015jva}.   
 \begin{figure}[H]
\centering\includegraphics[width=8.7cm]{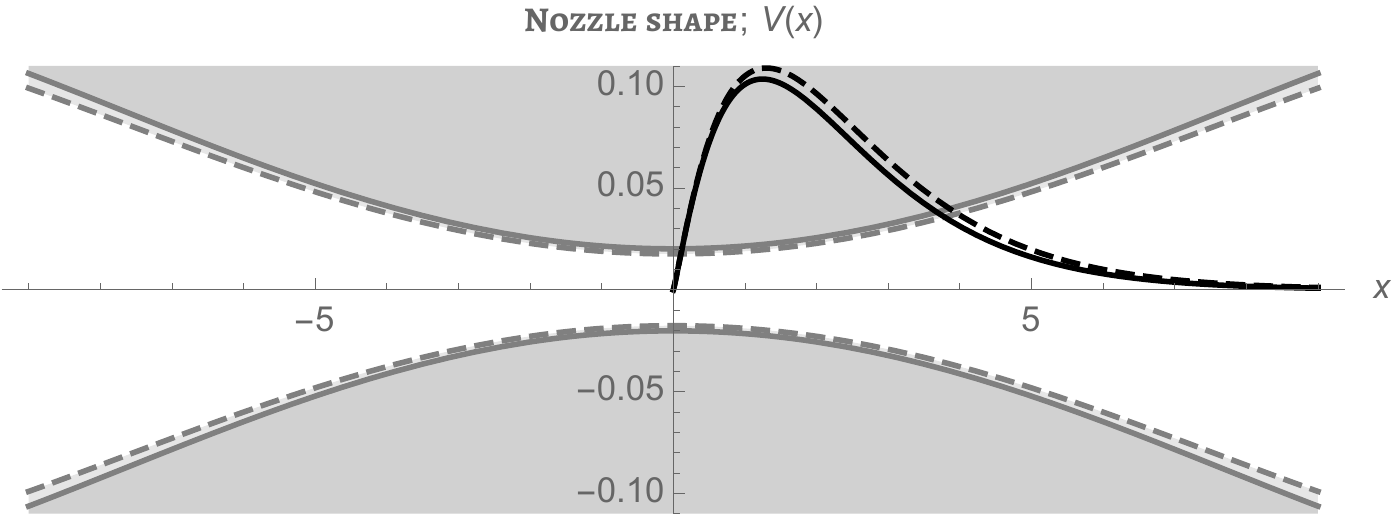}
\caption{The nozzle profile for $s=\ell=0$. The gray filled area, limited by the continuous [dashed] line denotes the de Laval nozzle in the Schwarzschild, GR $\sigma\to\infty$, limit [in the minimal geometric deformation of the de Laval nozzle]. The black lines represent $V(x)$ for the GR limit (continuous line) and its minimal geometric deformation (dashed line). }  \centering\includegraphics[width=8.7cm]{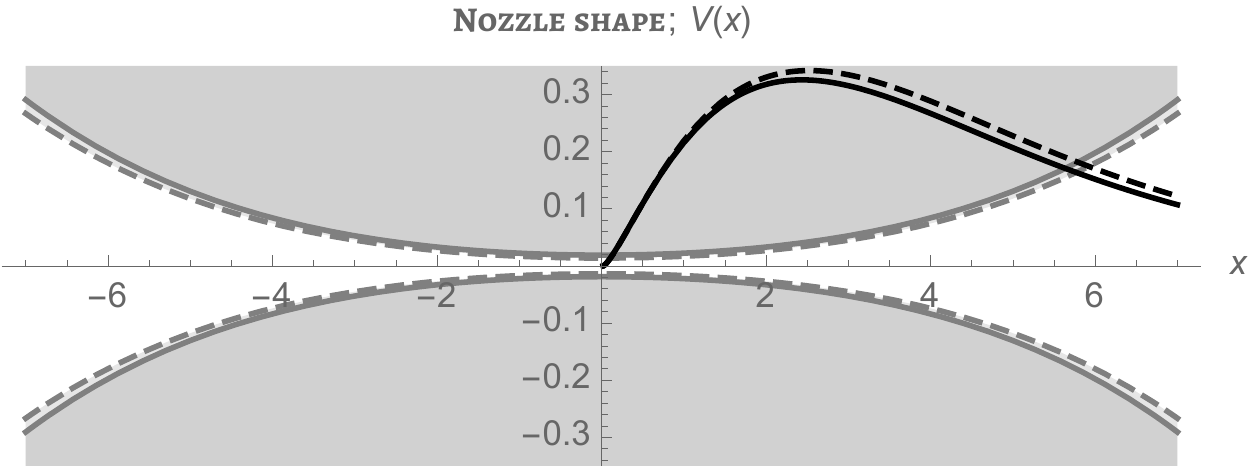}
\caption{The nozzle profile for $s=\ell=1$. The gray filled area, limited by the continuous [dashed] line denotes the de Laval nozzle in the Schwarzschild, GR $\sigma\to\infty$, limit [in the minimal geometric deformation of the de Laval nozzle]. The black lines represent $V(x)$ for the GR limit (continuous line) and its minimal geometric deformation (dashed line). }
\label{ppppp1} \centering\includegraphics[width=8.7cm]{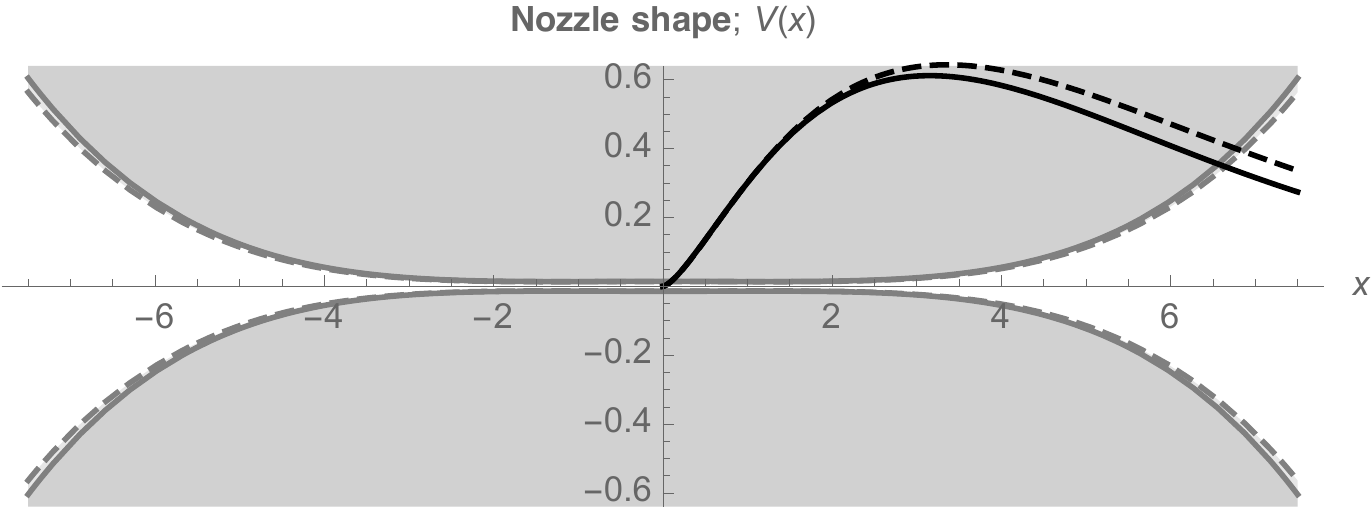}
\caption{The nozzle profile for $s=\ell=2$. The gray filled area, limited by the continuous [dashed] line denotes the de Laval nozzle in the Schwarzschild, GR $\sigma\to\infty$, limit [in the minimal geometric deformation of the de Laval nozzle]. The black lines represent $V(x)$ for the GR limit (continuous line) and its minimal geometric deformation (dashed line). }
\label{ppppp1}
\end{figure}
The minimal geometric deformation constricts the nozzle throat cross-sectional area. 
In addition, by comparing the Schwarzschild versus minimal geometric deformation, the nozzle corrections due to the influence of a 5D bulk Weyl fluid that permeates the vacuum on the brane are notorious. 
Hence,  the quasinormal modes of  black holes solutions deformed by the 5D bulk Weyl fluid can be probed by their analogy with acoustic perturbations of a diatomic gas fluid flow in a de Laval nozzle, using the geometric deformation technique. 
Reciprocally, the signature and the lacking data about the Weyl tensor in the bulk can be probed in a laboratory, by analysing sonic waves throughout a de minimal geometrically deformed de Laval nozzle. The next section is devoted to further explore and analyse the consequences of 
our results. 

\section{Concluding remarks and outlook}

                The minimal geometric deformation of a de Laval nozzle can have double-handed applications. The de Laval nozzle associated with black hole analogues produced in the laboratory can present their trend slightly modified by a 5D bulk Weyl fluid effects. On the other hand, 5D effects can be also probed in the laboratory, due to the analogy heretofore presented. 
                
                Using a phenomenological E\"otv\"os-Friedmann fluid brane setup, describing an inflationary brane-world universe, the perturbation of a fluid flow in a de Laval nozzle was considered,  
                providing a wave equation that is similar to the wave equation regarding perturbations of  minimal geometrically deformed brane-world black holes. The precise bounds for the variable brane tension value provided corrections to the shape of de Laval nozzles in this context. Figs. 1 - 3 plot the de Laval nozzle
                profile and also the analysis of the nozzle deformation with respect to the Schwarzschild solution. Such a deformation is  generated by a 5D Weyl fluid permeating a compact distribution described by the Schwarzschild metric solution of the 4D brane effective Einstein's equations.
                 
Moreover, quasinormal modes  of brane-world black holes undergoing a minimal geometric deformation can be, then, produced and observed in a laboratory, by analysing the sonic waves throughout the associated deformed Laval nozzle. Hence, the solution for the inverse technique, consisting of the correspondence between the shape of the de Laval nozzle and the general trend of perturbations 
in brane-world black holes deformed by a minimal geometric deformation of Schwarzschild black holes, has been here implemented. The corrections to the Schwarzschild solution on the brane, permeated by a 5D bulk Weyl fluid, affect how the pressure is dispersed across the deformed de Laval nozzle. 
The finite brane tension, then, specifies a protocol to the analysis consisting of whether the
de Laval nozzle highest thrust can be achieved and the search for best flow properties that are being attained, for the derived de Laval nozzle shape.

Using the sonic analogue to black holes \cite{Unruh:1994je}, the  thermal spectrum of sound waves was given out from the sonic horizon in transsonic fluid flows, also in the context of analogue gravity \cite{Visser:1997ux,Barcelo:2005fc,Crispino:2007zz}. These approaches can be further explored, using the methods here introduced, together with more fluid analogies phenomena regarding  black holes in the laboratory \cite{Schutzhold:2002rf,Mosna:2016nrt}. 
Still, further types of black holes can be studied \cite{Lin:2014vla, Morgan:2009vg}. Finally, the extended MGD approach \cite{Casadio:2015gea} can 
also be used to derive further corrections to the de Laval
nozzle profile.

\subsection*{Acknowledgements}

RdR~is grateful to CNPq (Grant No. 303293/2015-2),
 to FAPESP (Grant No.~2015/10270-0), for partial financial support, and to Dr. A. Zhidenko for worth discussions. 

\bibliography{bib_deLaval}

\end{document}